\documentclass[a4paper,10pt]{article}
\usepackage[utf8]{inputenc}
\usepackage{authblk}
\usepackage{amsmath}
\usepackage{graphicx}
\usepackage{cite}

\title{\Large \bf Stochastic resonance and optimal information transfer at criticality on a network model of the human connectome}
\author[1,*]{Bertha Vázquez-Rodríguez}
\author[2]{Andrea Avena-Koenigsberger}
\author[2]{Olaf Sporns}
\author[3,4]{Alessandra Griffa}
\author[3,4]{Patric Hagmann}
\author[1]{Hernán Larralde}
\affil[1]{\footnotesize{Instituto de Ciencias Físicas, Universidad Nacional Autónoma de México}}
\affil[2]{\footnotesize{Department of Psychological and Brain Sciences, Indiana University}}
\affil[3]{\footnotesize{Department of Radiology, Lausanne University Hospital}}
\affil[4]{\footnotesize{University of Lausanne}}

\affil[*]{bertha@fis.unam.mx}
\date{}
\begin{document}

\maketitle

\begin{abstract}

Stochastic resonance is a phenomenon in which noise enhances the response of a system to an input signal. The brain is an example of a system that has to detect and transmit signals in a noisy environment, suggesting that it is a good candidate to take advantage of SR. In this work, we aim to identify the optimal levels of noise that  promote signal transmission through a simple network model of the human brain. Specifically, using a dynamic model implemented on an anatomical brain network (connectome), we investigate the similarity between an input signal and a signal that has traveled across the network while the system is subject to different noise levels. We find that non-zero levels of noise enhance the similarity between the input signal and the signal that has traveled through the system. The optimal noise level is not unique; rather, there is a set of parameter values at which the information is transmitted with greater precision, this set corresponds to the parameter values that place the system in a critical regime. The multiplicity of critical points in our model allows it to adapt to different noise situations and remain at criticality.

\end{abstract}

\section*{Introduction}
Random noise has been traditionally considered as an obstacle in the transmission of information, contaminating accurate communication and limiting the achievable information rate \cite{weaver}. Nonetheless there are examples in which the presence of noise makes substantial improvements in signal detection \cite{douglass, mendez, cats}, through the phenomenon of stochastic resonance (SR).

SR was proposed as a possible explanation for the periodicity of the ice ages on Earth \cite{benzi}, and has been studied in Schmitt triggers\cite{fauve}, tunnel diodes\cite{mantegna} and bidirectional ring lasers\cite{mcnamara}. Nowadays, the effects of noise on biological sensory systems is being extensively explored. One of the first demonstrations of SR in the nervous system was carried out on crayfish mechanoreceptors \cite{douglass, srcircle}. Since then, other experimental demonstrations have included neurons in crickets\cite{cricket}, rats\cite{rats1,rats2}, and cats\cite{cats}, along with several studies in humans on the enhancement of detection and transmission in the sensorimotor system during a motor task \cite{mendez, collins}.
In \cite{hutt} the propagation of a periodic input signal through an Erdös-Rényi network for different noise levels was studied, and it was found that noise indeed enhanced signal propagation in the model. However, to our knowledge, no studies have explored the SR phenomenon as a mechanism that could potentially enhance the transmission of information along axonal pathways in the human brain.

Growing evidence supports the hypothesis that the dynamics of the brain resembles the dynamics of a system near a critical point. This suggests that many functionally important features of brain dynamics may be optimized at criticality \cite{chialvo1, chialvo2, beggs1, beggs2, shew}. Recent work has shown that a discrete state dynamical model implemented on a network of neuroanatomical connections (connectome\cite{olaf}) exhibits a phase transition similar to that observed in a percolation model, where the average size of the second biggest cluster of active nodes reaches its maximum value for a specific activation threshold (\cite{haimovici}). Furthermore, the model presented in \cite{haimovici} is capable of replicating spontaneous brain activity patterns that resemble so-called resting state networks \cite{yeo}, which are widely regarded as key components of functional brain architecture \cite{buckner}. Other experiments have demonstrated that the dynamic range (the range of stimulus intensities that allows network responses to be distinguished) \cite{kinouchi}, mutual information and information capacity appear to be maximized at critical points \cite{shew2013,larremore}. It is not clear, however, whether the levels of noise that increase the transmission of information through the brain are related to criticality.

This issue is important because the brain, even when noise sources are present, must be capable of  integrating  information across multiple sensory modalities and brain systems, in order to generate adaptive neural and behavioral responses. In the present work we propose a dynamic model to determine quantitatively the amount of noise required for the best transmission of signals through the structural network of the brain's connectome and its relationship with criticality.

\section*{Results}

\subsection*{The model}

The model is implemented on a network representing a human connectome. Each node in this network represents a gray matter region of the human cortex whereas edges represent white matter fiber tracts that connect cortical regions. The weighted elements ($w_{ij}$) of the adjacency matrix of the connectome (figure \ref{fig.connectome} in the \textit{Methods section}) are proportional to the number of streamlines connecting two brain regions, indicating the strength of a connection between nodes $i$ and $j$. The method used to obtain the weighted network matrix of the human connectome is reported in \cite{hagmann} and is briefly described in the Supplementary Information section. 

The network contains $N=114$ nodes with binary states that are updated synchronously according to a dynamic rule adapted from \cite{haimovici}. Each node, characterized by a boolean variable $s_i$, is updated every time step and can be in one of two states: quiescent $Q$ (with $s_i=0$) or excited $E$ (with $s_i=1$). 
The state of each node obeys the following transition rules:\\
\begin{itemize}
 \item $Q\rightarrow E$ with a probability $P_{QE}$ (corresponding to spontaneous activation of the node) or if the input signal $\alpha_i=\sum\limits_{j=0}^{k_i}w_{ij}s_j(t)$ is higher than a threshold $T$.
 \item $E\rightarrow E$ with a probability $P_{EE}$ if the node was still stimulated to be activated as above, i.e. with a probability $P_{QE}$ or if $\alpha_i>T$.
 \item $E\rightarrow Q$ with a probability $(1-P_{EE})$ or with a probability $P_{EE}$ provided that the stimulus received is not large enough to maintain the node active.
\end{itemize}

The probability $P_{QE}$, i.e. the probability of spontaneous activation, plays the role of the noise in the system, thus $P_{QE}$ is the quantity that we expect to be connected with SR. $P_{EE}$ represents the probability that the node has enough material to fire for more than one time step (as may be the case if some of the neurons in that brain region have not fired yet). It is important to highlight that there is no refractory state in the model because the nodes represent whole brain regions comprising large populations of neurons, not individual nerve cells.

Thus the state of the i-th node changes in time according to the following dynamical rule:
\begin{align}
 s_i(t+1)=&\left\{ 1 + s_i(t)[H(P_{EE}-r_2)-1]\right\}\times \nonumber\\ 
 &\left\{H(P_{QE}-r_1)+\left[ 1 - H(P_{QE}-r_1) \right] H(\alpha_i-T)\right\} \nonumber
\end{align}
\noindent where $r_1$ and $r_2$ are independent random numbers drawn from a uniform distribution between zero and one; and $H(x)$ is a step function (with $H(x) = 1$ if $x\geq 0$ and $H(x) = 0$ otherwise).

\subsection*{Phase space and criticality}

To gain a deeper understanding of the dynamics of our system we determined the phase space considering the time average $\langle S\rangle$ of the instantaneous mean activity $S(t) = \frac{1}{N} \sum\limits_{i = 0}^N s_i(t)$ as the order parameter --specifically  $\langle S\rangle = \frac{1}{L}\sum\limits_{t=t_0}^{t_0+L} S(t)$  (where $t_0$ is a transient time in which the system equilibrates)-- and $P_{QE}$, $P_{EE}$ and $T$ are the control parameters.

\begin{figure}[ht!]
 \centering
\includegraphics[width=.8\linewidth]{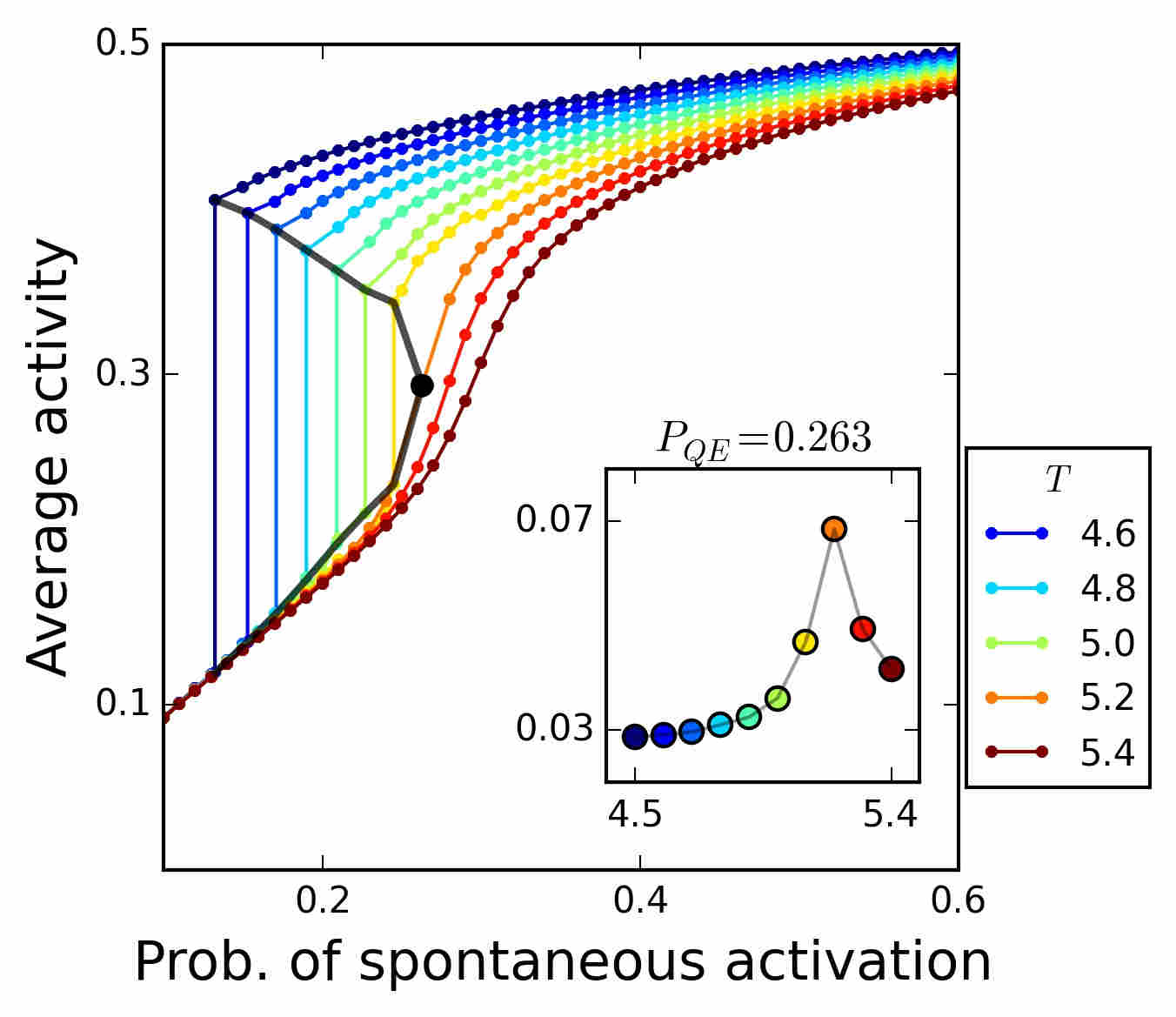}
\caption{\footnotesize{Average activity as a function of $T$ and $P_{QE}$ for a fixed value of $P_{EE}=0.1$. When the system is in the region inside the coexistence curve, the activity will be jumping around two different values. Outside this region the level of activity will be fluctuating around one single value. The black dot represents the critical point for this level of $P_{EE}$. Inset: standard deviation of the average activity for $P_{QE}=0.263$ and all the values of $T$.}}
\label{fig.coex}
\end{figure}

\begin{figure}[ht!]
 \centering
\includegraphics[width=\linewidth]{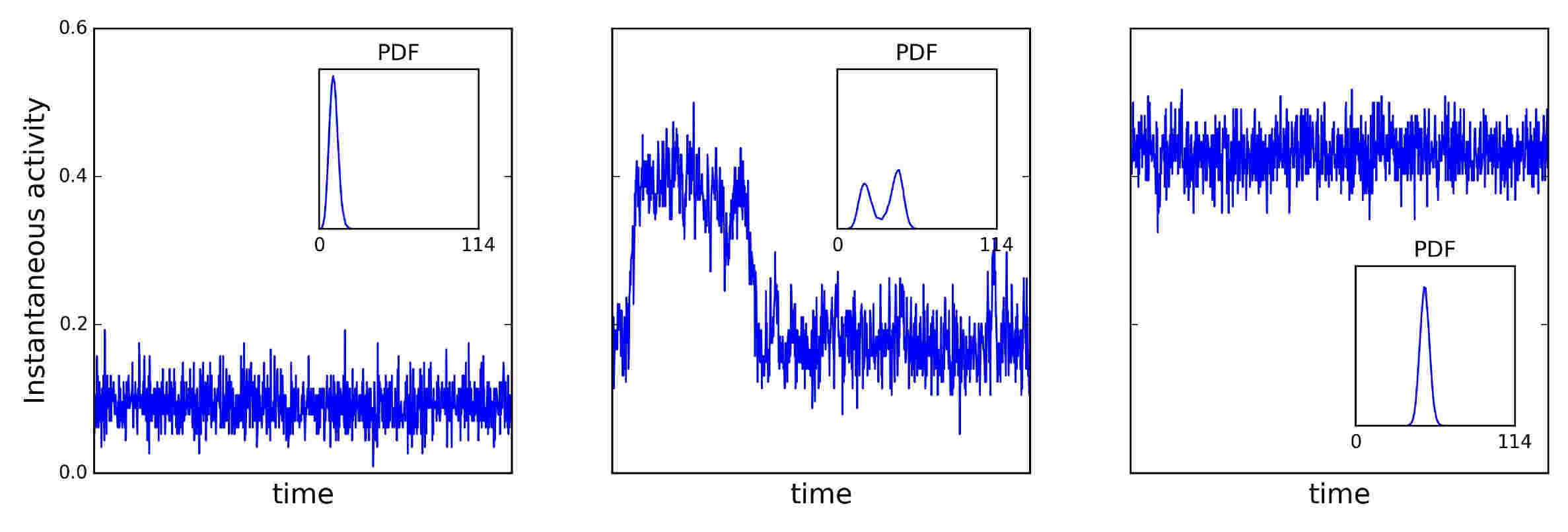}
\caption{\footnotesize{Instantaneous mean activity for a system with $P_{EE}=0.1, T = 4.8$ and $P_{QE}=0.1, 0.19$ and $0.3$. Inset: the probability density function for the number of active nodes. For the first $P_{QE}$ the system is on a low activity level. When it is near the coexistence curve $S(t)$ jumps between the high and low activity levels. For the largest $P_{QE}$ value, the system is on the high activity phase.}}
\label{fig.Sstd}
\end{figure}

Figure \ref{fig.coex} shows the phase space for a fixed value of $P_{EE}$ and varying values of $T$ and $P_{QE}$. For a large value of $T$ the average activity is a monotonically increasing function of $P_{QE}$. For small values of both $T$ and $P_{QE}$, the system's average activity is initially low, but as $P_{QE}$ increases, the system varies discontinuously from a low activity phase to a high activity phase. Hence, there is a range of $P_{QE}$ values for which the activity of the system will jump between the high and low activity phases (figure \ref{fig.Sstd}, middle panel). In order to determine the average activity $\langle S \rangle$ associated to the low and high activity phases taking place at a fixed value of $P_{QE}$, we notice that the PDF of $S(t)$ is bimodal when the system is unstable and jumping between a high and low activity phase (see figure \ref{fig.Sstd}, inset in middle panel). Hence, we fit two Gaussian distributions to each mode of the PDF of $S(t)$; each of the mean of the two Gaussian distributions represent the average activity $\langle S \rangle$ of the low and high activity phases, respectively. The standard deviation of the activity is then computed as the mean of the standard deviations of each of the two Gaussian distributions. 

For a fixed $P_{EE}$, the corresponding {\it coexistence curve} is defined by all the points in the phase space such that the PDF of $S(t)$ is bimodal. The black line in Figure \ref{fig.coex} shows the coexistence curve for $P_{EE} = 0.1$. Notice that, within the coexistence curve, as T increases, the difference between the average high and low activity phases decreases, which is the result of the two modes of the PDF of $S(t)$  approaching each other. Eventually, the two modes converge onto a single mode PDF and the average activity of the system fluctuates around a single value (black marker in Figure \ref{fig.coex}). Interestingly, at this point, the standard deviation of the system's activity  $S(t)$ reaches its maximum, as shown in the inset in figure \ref{fig.coex}. At this point, the system is {\it critical}.

\begin{figure}[ht!]
 \centering
\includegraphics[width=.9\linewidth]{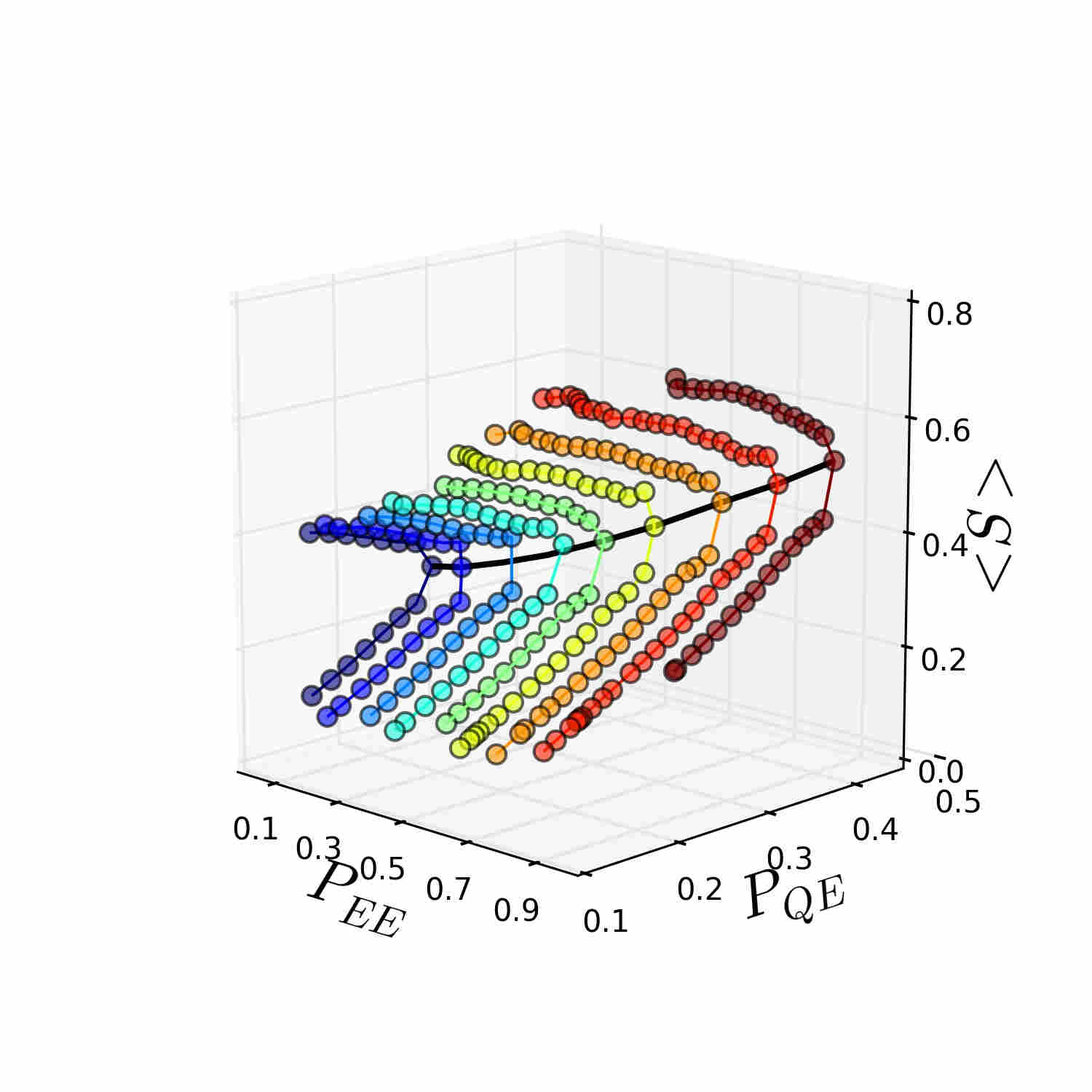}
\caption{\footnotesize{Coexistence curves. The colors correspond to different $P_{EE}$ values for a network of $N = 114$ nodes. It can be seen that there is a set of parameters that tune the system into a critical state (black curve).}}
\label{fig.vdw}
\end{figure}

For each value of $P_{EE}$ we find the corresponding coexistence curve and the critical point within the phase space. The set of coexistence curves are shown in Figure \ref{fig.vdw}. Within the parameter space ($\langle S \rangle$,$P_{QE}$,$P_{EE}$), the set of coexistence curves delimits the region where the system exhibits the unstable high and low activity phases. Furthermore, the coexistence curves end at the critical points (black line, fig. \ref{fig.coex}). 

\subsection*{Capturing the transmission of signals through the network}

To evaluate the transmission of information through the network, we introduced a signal in the system, one node at a time. This signal consisted in switching the node's activity to be in the {\it on} state for a certain number of time steps and then in the {\it off} state for the same number of steps; this pattern was continued periodically throughout the evolution of the system. While the input signal was being delivered through a specific input node $i$, we let the rest of the system evolve according to the dynamic rule and evaluate the {\it similarity} between the input signal and the output signal -the activity- at each node. 

We use the Fourier spectrum of the signals to measure their {\it similarity}. If the difference between two signals is that they are merely rescaled, their power spectrum will have the same principal frequencies, but not the same amplitudes. Yet, we would want to say that the signals are {\it similar}. Accordingly, we search for a factor ($\lambda_{ij}$) that will minimize the weighted squared difference between the input and output spectrum. To determine this factor, each term of the squared difference is multiplied by the amplitude of the input spectrum, so the frequencies with the larger amplitudes will have a greater contribution in the sum, and hence, in the determination of $\lambda_{ij}$.  Finally, we define the similarity between the signals as:

\begin{eqnarray}
sim(i,j)= -\log\left( \frac{1}{\sum\limits_m^M (\phi_m^i)^3} \sum\limits_n^M (\phi_n^{i} - \lambda_{ij}\phi_n^{j})^2 \phi_n^{i} \right),\nonumber 
\end{eqnarray}
wherea
\begin{eqnarray}\lambda_{ij}=\frac{\sum_m^M (\phi_m^{i})^2\phi_m^{j}}{\sum_m^M (\phi_m^j)^2 \phi_m^{i}} \nonumber
\end{eqnarray}
and $\phi_n^i$ is the amplitude of the n-th main frequency of the signal input in node $i$ (input node, or "seeder"), and $\phi_n^j$ the amplitude for the same frequency in the power spectrum of the activity at node $j$ (output node, or "receiver"). To compute this quantity we use only the $M$ principal frequencies of the node $i$, that are the ones that have an amplitude larger than $0.0001$. The normalization factor allows us to compare this measure for different input signals and the $-log$ will make $sim(i,j)$ maximum when the difference between the two spectra is the lowest. 

When the system is near the coexistence curve it will be jumping between two levels of activity. This behavior could affect the measure of similarity, so we did not select states contained in the coexistence curve. We calculated the {\it similarity} for different sets of parameters within the line of critical points (black line in figure 3) and non critical states that are not in the coexistence regions. For the remainder of the manuscript we show results corresponding to  $P_{EE} = 0.1$, but results are qualitatively similar for other values of $P_{EE}$ (see figure \ref{fig.05_09} at the \textit{Methods section} for figures corresponding to $P_{EE} = 0.5$ and $0.9$).

\begin{figure}[ht!]
  \centering
  \includegraphics[width=\linewidth]{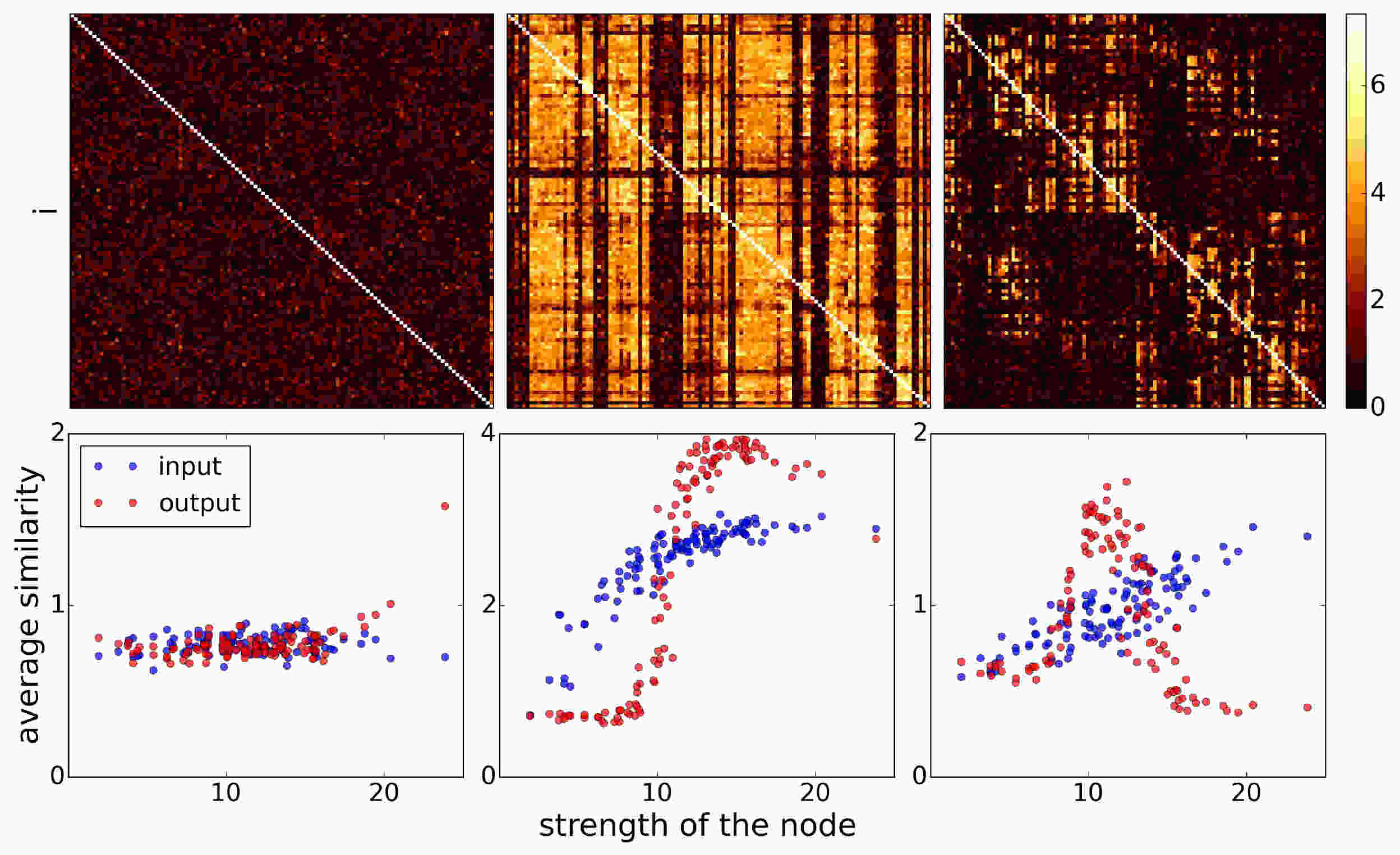}
  \caption{\footnotesize{Top: {\it Similarity} matrices evaluated for parameters  $P_{EE}=0.1$, $T=5.2$ and $P_{QE}=0.15$ (non critical with low activity level), $0.263$ (critical) and $0.4$ (non critical with high activity level). Bottom: Average similarity over inputs (blue) or outputs (red) as a function of the strength of the nodes $w_i=\sum_j^{k_i} w_{ij}$.}}
  \label{fig.simi}
\end{figure}

Figure \ref{fig.simi} (upper panel) shows examples of pairwise similarity matrices evaluated at two non-critical points (left and right panels) and a critical point (center panel). The $i^{th}$ row of these matrices corresponds to an instance in which the input signal was introduced through node $i$ and all other nodes act as output nodes for which we measure the {\it similarity} between their activity time-series and the signal introduced at node $i$. We note that for these experiments we use the same input signal for all nodes.

Our results show that similarity values are significantly higher for most pairs of input-output nodes when the system is critical (figure \ref{fig.simi}, upper middle panel). In other words, when the system is critical, the noise level, implemented by the parameter $P_{QE}$, facilitates the transmission of an input signal across most neural elements of the network  (see figure \ref{fig.all_simi} at the \textit{Methods section} for a different set of critical parameters). However, the input signal is not transmitted to the entire system: we find that there is a set of ``deaf'' nodes whose output signal shows extremely low similarity to the input signal, regardless of what node we select to introduce the input signal. This behavior is expressed by the dark column-like patterns in the similarity matrix of the critical point. Interestingly, these nodes tend to have higher values of similarity when the system is not critical. Thus, the input signal's principal frequencies are suppressed at these nodes when the system is critical. It is also worth noting that the column-like patterns expressed in the similarity matrices at the critical point suggest that, with a few exceptions (e.g. output-nodes within the right hemisphere frontal pole, right hemisphere medial orbitofrontal cortex, right hemisphere parahippocampal region, right hemisphere entorhinal region, right hemisphere temporal pole, left hemisphere postcentral region, left hemisphere supramarginal gyrus, and left hemisphere transversal temporal gyrus), the system's dynamics at criticality do not vary greatly as a result of varying the input node. This is not the case when the system is not critical in which case the patterns of similarity vary depending on who the input node is, as shown by the variability across rows in the similarity matrices shown in figure \ref{fig.simi}, upper right and left panels. 

In order to gain more insight about the identity of these deaf nodes and what causes these nodes to be deaf to the input signal when the system is critical, we examined the average similarity of each output node across all input nodes (i.e. we compute an average across the columns of the similarity matrix) and the average similarity of each input node across all outputs (i.e. we compute an average over the rows of the similarity matrix).

\begin{figure}[ht!]
  \centering
  \includegraphics[width=.8\linewidth]{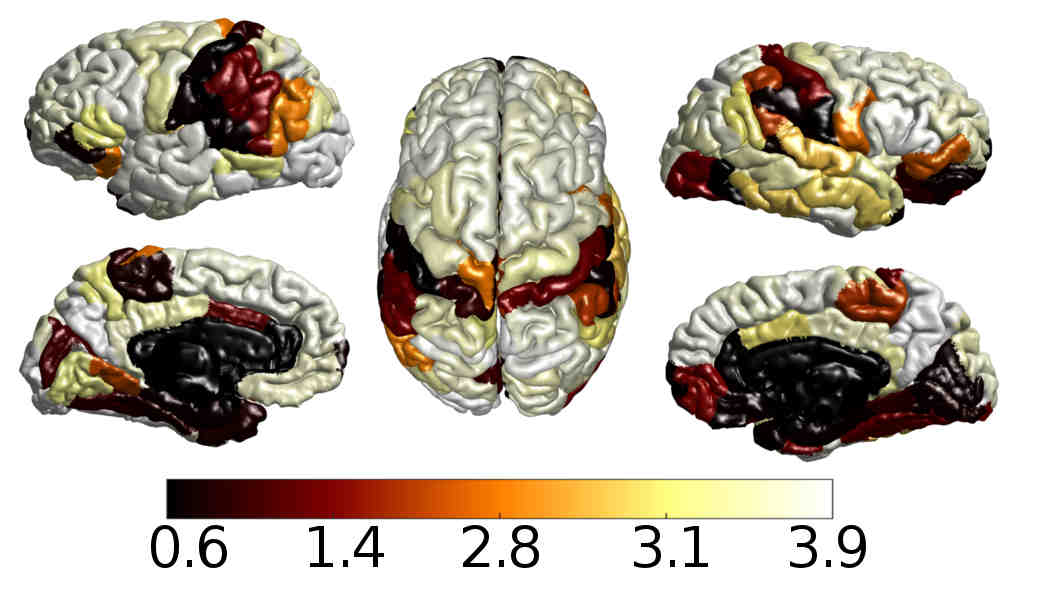}
  \caption{\footnotesize{Average similarity across output nodes projected on the human cortical surface. Anatomical brain regions that suppress the input signal's principal frequencies are: bank of the superior temporal suculus, cuneus, entorhinal cortex, frontal pole, fusiform gyrus, lateral occipital cortex, lateral orbitofrontal cortex, lingual gyrus, medial orbitofrontal cortex, paracentral lobule, parahippocampal cortex, pars orbitalis, postcentral gyrus, rostral anterior cingulate cortex, superior temporal cortex, supramarginal gyrus, temporal pole, transverse temporal cortex.}} 
  \label{fig.surf_plot}
\end{figure}

Figure \ref{fig.simi} (bottom panel) shows similarity averages across inputs (blue markers) and outputs (red markers) as a function of the strength of the nodes where strength is defined as the total sum of all connection weights of a node. We note that at the critical point (lower middle panel) and at the non-critical point with high average activity $\langle S \rangle$ (lower right panel) we find a clear relationship between average similarity values and node strength. Yet, the relationship between average similarities and node strength varies. At the critical point, the average similarity across output nodes resembles a step function with input-output similarity  drastically increasing when the node strength exceeds a threshold value. Figure \ref{fig.surf_plot} shows the average similarities across outputs projected on a template cortical surface, allowing us to identify the location of ``deaf nodes'' (indicated by dark colors). Outside of criticality the output average similarity increases with node strength at first, but then decreases as node strength continues to increase. Average similarity across input nodes increases slowly as a function of node strength. For a non-critical point with low average activity (left panel), we find no relationship between average similarity and node strength. 
In the supplementary information section we show qualitatively similar results obtained from other measures used to assess the similarity between input and output signals, such as mutual information (figures \ref{fig.mi}, \ref{fig.mi_iTau} at the \textit{Methods section}) and dynamic correlation (figures\ref{fig.corr}, \ref{fig.corr_iTau} at the \textit{Methods section}) between nodes. 

\section*{Discussion}

The main goal of this work was to determine whether noise can possibly enhance the transfer of information within a simple dynamic model of the brain, and if so, to determine whether this noise corresponds to the value that tunes the system to a critical state. Our findings indicate that a noise level different from zero indeed promotes signal transmission and communication through the network, in line with what experimental evidence has shown\cite{mendez,douglass}. 

Further, we confirm that when the system is in a critical state, transmission of signals aided by noise is optimal. Additionally, having explored the system's entire phase space (figure \ref{fig.vdw}) we have found a set of parameter values at which the system is at criticality. This contrasts with previous models, which are critical at a unique point in their parameter space\cite{haimovici}. The multiplicity of critical points in our model may be behind the brain's capacity to adapt to different situations. Thus, for example, if the intensity of the noise changes, the system may adjust its parameters to remain at criticality. 

Hütt et. al. have found in \cite{hutt} that the number of outputs excited by the propagation of a periodic signal through their network is maximum for a noise level different from zero. One important difference with our work is that we are not only interested in the propagation of the signal up to the outputs, but also in  preservation of the signal shape. 

We stress that even when SR could be observed at different noise levels, the critical regime appears as the best condition for the transfer of signals through the system. Measurements of similarity revealed that the values obtained in the critical regime are larger than those measured when the system is at a non-critical state (\ref{fig.simi}). Interestingly, even when the system is at a critical state, there is a set of nodes that exhibit an incapacity to communicate with the rest of the network. This set can be seen as dark columns with values near to zero in the center of figure \ref{fig.simi}.

The system studied in this paper is limited by the small size of the connectome network we used (114 nodes), making it difficult to find the exact critical point. These networks were extracted from the combination of diffusion spectrum imaging and tractography, a widely used approach for non-invasive reconstruction of human anatomical connectivity. In future work, new non-invasive technologies are likely to contribute more detailed maps of anatomical brain networks in humans. An intriguing avenue for further investigation would be to examine individual differences in signal transmission, and changes across development and life span.

Given the simplicity of the model presented here, future research could aim at finding the differences in dynamical properties using the same analysis over networks extracted from diseased brains. If the transmission of information in damaged networks is different, the noise effect described in our work could be explored as a way to improve neural communication.  Another goal could be to study the cooperative and competitive effects in the spreading of a signal through neural networks \cite{bratis}. It would also be interesting to study how these effects change according to the seeder nodes. These and other extensions of the present study could assist to our understanding of how communication processes contribute to various aspects of brain function.

\section*{Methods}

\subsection*{Data}

Informed written consent in accordance with the Institutional guidelines (protocol approved by the Ethics Committee of Clinical Research of the Faculty of Biology and Medicine, University of Lausanne, Switzerland) was obtained for all subjects. Forty healthy subjects (16 females; 25.3 $\pm$ 4.9 years old) underwent an MRI session on a 3T Siemens Trio scanner with a 32-channel head coil. Magnetization prepared rapid acquisition with gradient echo (MPRAGE) sequence was 1-mm in-plane resolution and 1.2-mm slice thickness. DSI sequence included 128 diffusion weighted volumes + 1 reference b\_0 volume, maximum b-value 8,000 $s/mm^2$, and 2.2 $\times$ 2.2 $\times$ 3.0 mm voxel size. EPI sequence was 3.3-mm in-plane resolution and 3.3-mm slice thickness with TR 1,920 ms. DSI and MPRAGE data were processed using the Connectome Mapping Toolkit \cite{daducci2012connectome}. Each participant's gray and white matter compartments were segmented from the MPRAGE volume. The gray matter volume was subdivided into 68 cortical and 15 subcortical anatomical regions, according to the Desikan-Killiany atlas \cite{desikan2006automated}, defining 83 anatomical regions. These regions were hierarchically subdivided to obtain five parcellations, corresponding to five different scales \cite{cammoun2012mapping}. The present study uses a parcellation comprising 129 regions of interest (ROI); however, here we focus on cortical structures only, discarding all subcortical regions including the bilateral thalamus, caudate, putamen, pallidum, nucleus accumbens, hippocampus, and amygdala, as well as the brainstem, resulting in 114 remaining ROI. Whole brain streamline tractography was performed on reconstructed DSI data \cite{wedeen2008diffusion}, and connectivity matrices were estimated from the streamlines connecting each pair of cortical ROI. We quantify the connection strength between each pair of regions as a fiber density \cite{hagmann2008mapping} instead of fiber count. Thus, the connection weight between the pair of brain regions $\{u,v\}$ captures the average number of connections per unit surface between $u$ and $v$, corrected by the length of the fibers connecting such brain regions. The aim of these corrections is to control for the variability in cortical region size and the linear bias toward longer fibers introduced by the tractography algorithm. Fiber densities were used to construct subject-wise structural connectivity matrices. Finally, we construct a group connectome from all 40 individual subject connectomes following the consensus approach described in \cite{bratis}, where edges that are most frequently found across all individual are selected to conform the group connectome. Following the edge-weight transformation procedure described in \cite{hagmann}, our average connection matrix was obtained after fiber-density edge weights were re-sampled to a Gaussian distribution.

\begin{figure}[ht!]
 \centering
\includegraphics[width=.8\linewidth]{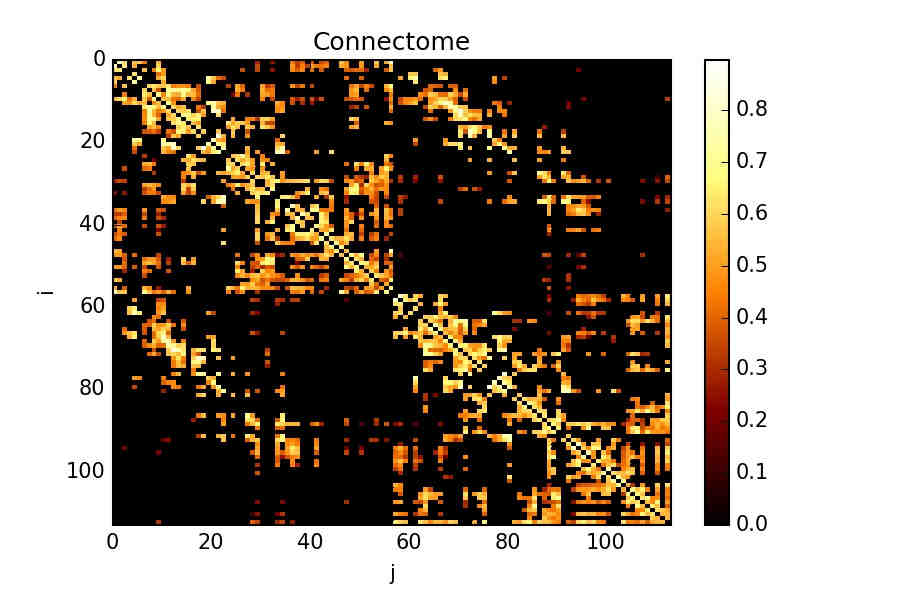}
\caption{\footnotesize{Representation of the adjacency matrix of the connectome for a network of 114 nodes. The colors in the matrix corresponded to the value of the connection between pairs of nodes. Nodes from 0 to 57 were located at the right hemisphere, whereas nodes from 58 to 114 were at the left hemisphere \cite{hagmann,olaf}. }}
\label{fig.connectome}
\end{figure} 

\subsection*{Phase space}

Phase space for two different levels of $P_EE = 0.5$ and $0.9$. The phase space is qualitatively similar to the one shown at the results section for $P_{EE}=0.1$.

\begin{figure}[ht!]
 \centering
\includegraphics[width=\linewidth]{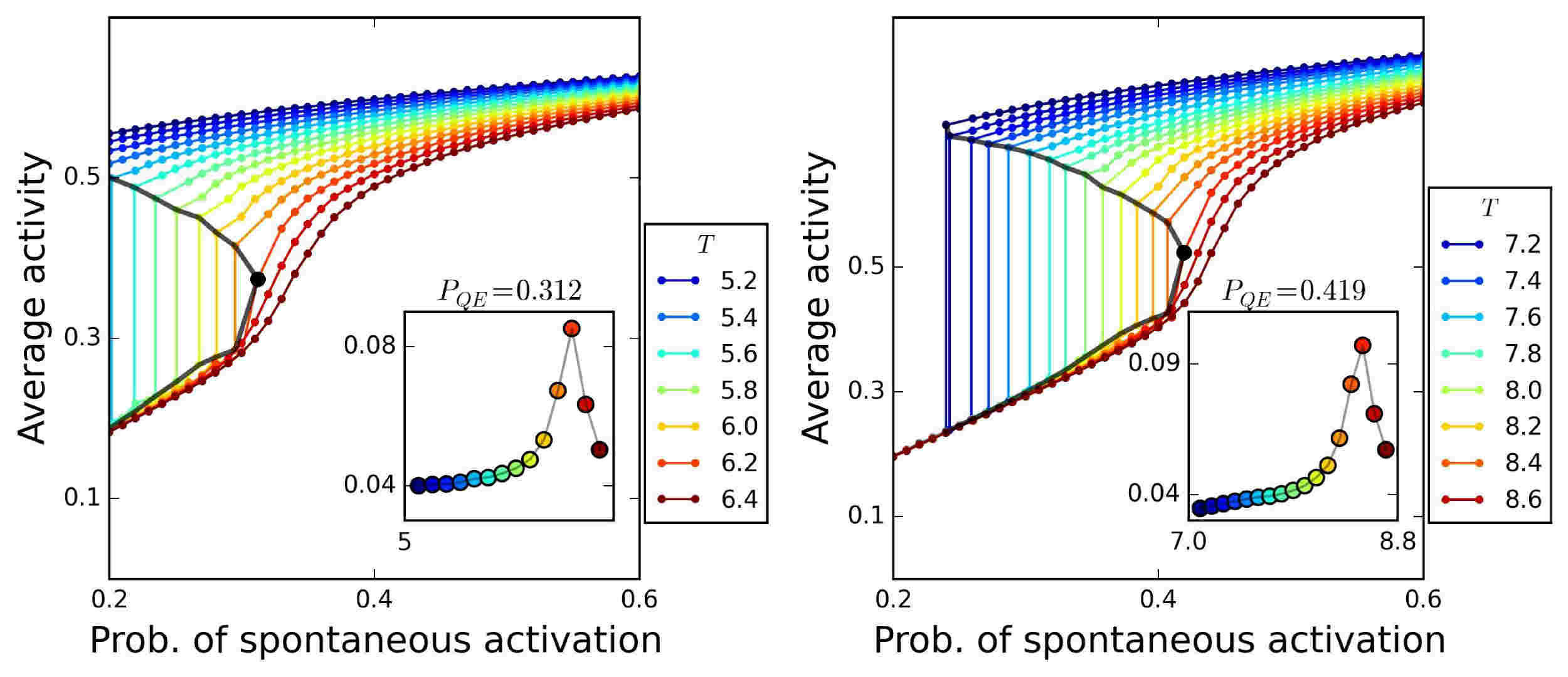}
\caption{\footnotesize{Average activity as a function of $T$ and $P_{QE}$ for a fixed value of $P_{EE}=0.5$ (left), $0.9$ (right). The black dot represent the critical point for this level of $P_{EE}$. Inset: standard deviation of the average activity for the same $P_{QE}$ and all the values of $T$.}}
\label{fig.05_09}
\end{figure} 

Similarity matrices evaluated at the corresponding critical point for three different levels of $P_{EE} = 0.1, 0.5$ and $0.9$. As we can see, the results at critical point does not depend on the probability of being on for more than one consecutive time step $P_{EE}$.

\begin{figure}[ht!]
  \centering
  \includegraphics[width=\linewidth]{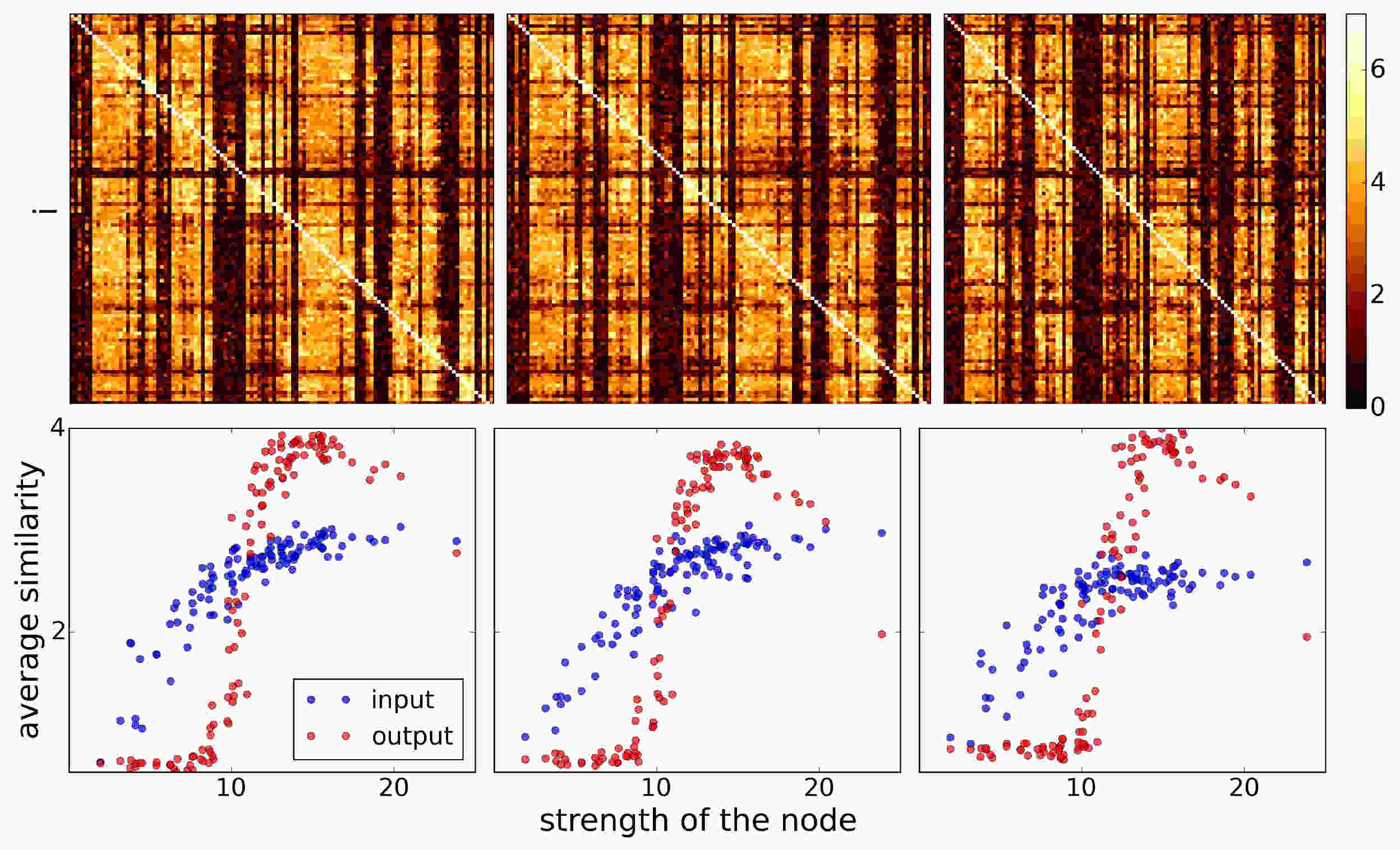}
  \caption{\footnotesize{Top: {\it Similarity} matrices evaluated for the critical parameters  for $P_{EE}=0.1, 0.5$ and $0.9$. Bottom: Average similarity over inputs (blue) or outputs (red) as a function of the strength of the nodes.}}
  \label{fig.all_simi}
\end{figure}

Overall, the phase diagram of our system (fig. \ref{fig.coex}) resembles to the liquid-gas transition described by the Van der Waals equation. In the Van der Waals theory of the liquid-gas transition, the isotherms can take different shapes on the pressure-density plane depending on the value of the temperature T. For a large value of T, the density is a continuous monotonically increasing function of the pressure. In contrast, when T is low enough, there are some values of the pressure in which the system can have two different densities. At these points the system undergoes a discontinuous phase transition from gas to  liquid or vice versa. The densities at which this transition occurs delimit the coexistence curve, where liquid and gas phases can coexist at the same temperature and pressure. The end point of the coexistence curve, at which the transition becomes continuous, is the critical point of the system, and lies on the isotherm corresponding to the critical temperature \cite{vdw}.

\begin{figure}[!ht]
 \centering
\includegraphics[width=.8\linewidth]{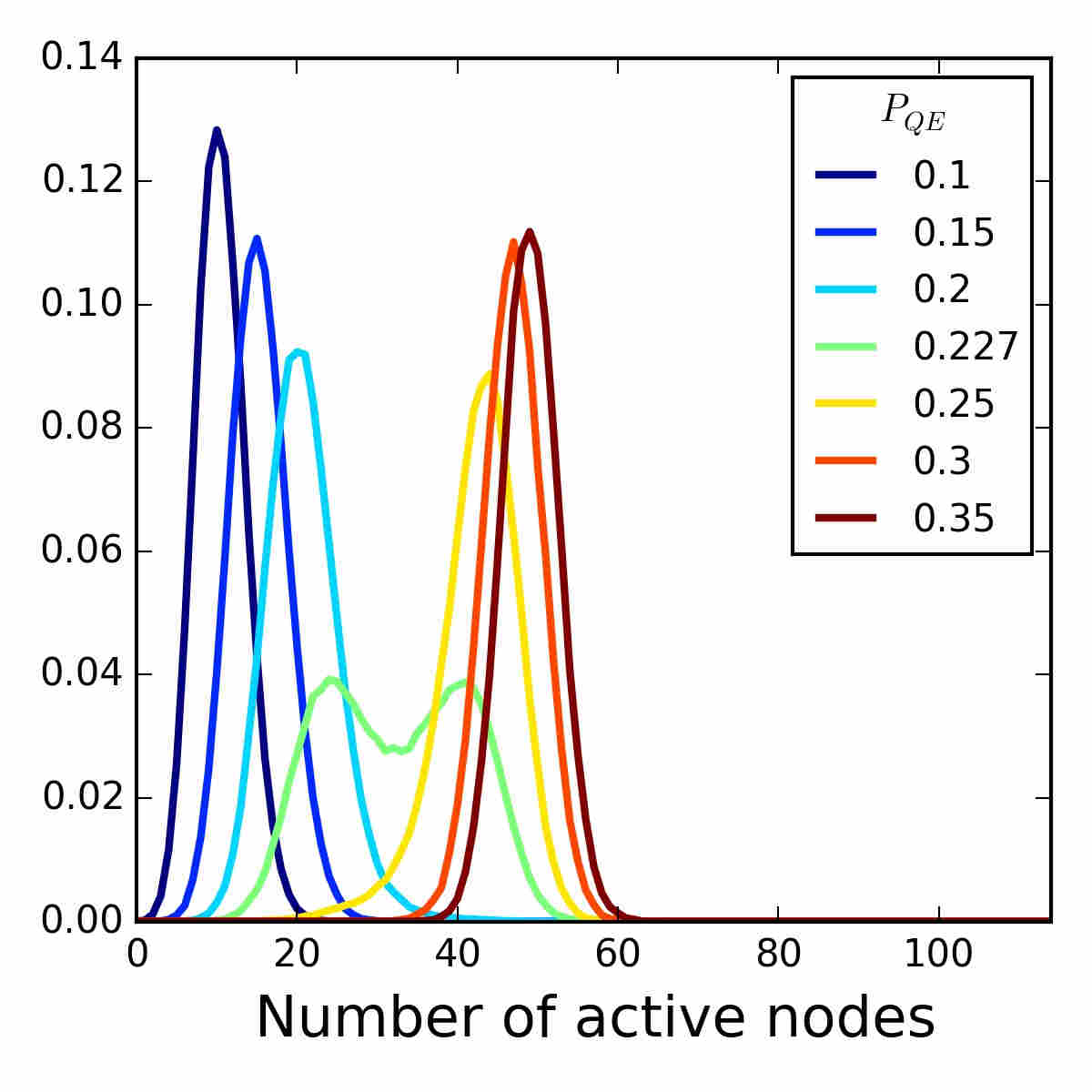}
\caption{\footnotesize{Probability distribution function for the number of active nodes for a system with $P_{EE}=0.1$ and $T=5.0$. As we increase $P_{QE}$ the system goes from a low to a high activity level. When the probability is bimodal ($P_{QE}=0.227$) the system is at the coexistence curve.}}
\label{fig.PDF_pQE}
\end{figure}

Making an analogy to the Van der Waals fluid, we constructed the phase space of our system in the following way. In order to determine the coexistence curve, we measured the distributions of values $S(t)$ attained in each run for fixed values of $P_{EE}$ and $T$, increasing $P_{QE}$ until the probabilities of being in the low or high activity state were the same (when the two maximums in the distribution of the activity have the same heights, as in figure \ref{fig.PDF_pQE}). We fit a sum of two Gaussian distributions to the PDF and the two "means" set the values of the average activity over the coexistence curve. Outside the coexistence curve, the values of the activity are simply given by the average activity value $<S>$.

\begin{figure}[!ht]
 \centering
\includegraphics[width=.8\linewidth]{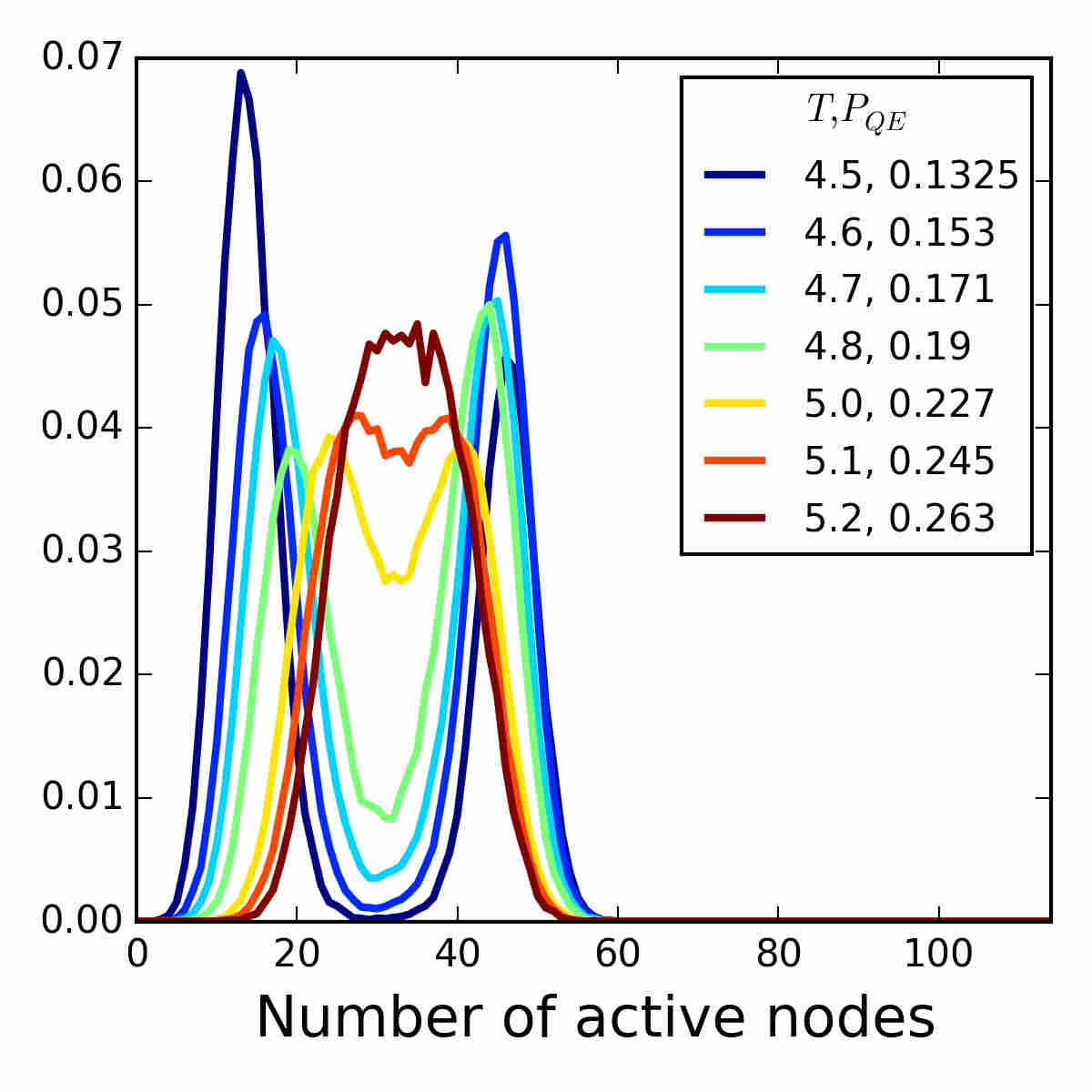}
\caption{\footnotesize{Probability distribution function for the number of active nodes for a system with $P_{EE}=0.1$. As we increase $T$ the difference between the low and high levels of activity decreases until they overlap at the critical value of $T$.}}
\label{fig.PDF_T_pQE}
\end{figure}

\begin{figure}[ht!]
 \centering
\includegraphics[width=\linewidth]{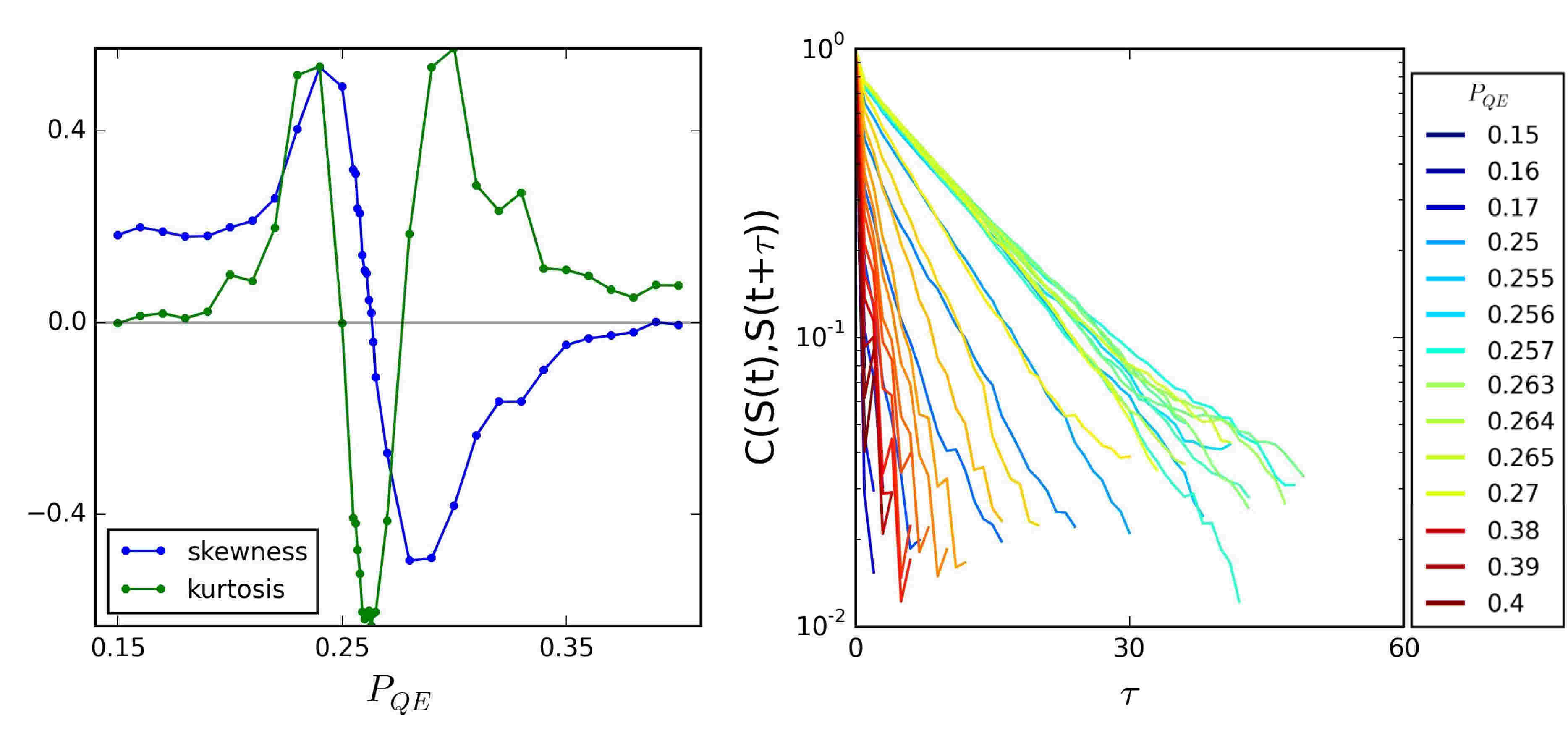}
\caption{\footnotesize{For a system with $P_{EE}=0.1$ and $T=5.2$ the skewness, kurtosis (left panel) and autocorrelation length (right panel) for different values of $P_{QE}$. We can see that when $P_{QE}$ is around 0.26 the kurtosis and autocorrelation length do not change too much and the skewness crosses zero, as we expect for the system when it is near criticality.}}
\label{fig.sk_kur_corr}
\end{figure}

When we increase the value of $T$, the difference between the two levels of activity at coexistence will decrease (the two maximums in the distribution will get closer). Thus, for a fix value of $P_{EE}$ we look for the $T$ at which the two levels of activity overlap (fig. \ref{fig.PDF_T_pQE}). Once we have the critical $T$, we compute the skewness and the kurtosis for the distributions, as well as the autocorrelation time obtained from runs at different values of $P_{QE}$. The distribution with a skewness closest to zero, a negative kurtosis and the largest autocorrelation time was chosen as critical (fig. \ref{fig.sk_kur_corr}). The activity of the system at this point will be fluctuating around a single value and will have the highest standard deviation.

\subsection*{Information measurements}

Other ways to quantify the transmission of information between nodes would be to measure mutual information and cross correlation as a function of the delay $\tau$. These measurements assess how the statistics of the activity at one node are related to the activity at another node.  
The mutual information between nodes $i$ and $j$ is defined as \cite{dayan}:

\begin{align}
 MI(s_i,s_j,\tau)=&\sum_{a=0,1}\sum_{b=0,1} P(s_i(t)=a,s_j(t+\tau)=b)\times \nonumber\\ 
 &\log\left( \frac{P(s_i(t)=a,s_j(t+\tau)=b)}{P(s_i(t)=a)P(s_j(t+\tau)=b)} \right), \nonumber
\end{align}
where $P(x)$ is the probability of $x$, and $P(x,y)$ is the joint probability for $x$ and $y$. 

Cross correlation with delay $\tau$ was computed as:
\begin{equation}
C(s_i,s_j,\tau)= \left|\frac{1}{(L-\tau)\sigma_i\sigma_j}\sum\limits_{t=0}^{L-\tau} (s_i(t)-\mu_i)(s_j(t+\tau)-\mu_j)\right|, \nonumber
\end{equation}
where $\mu_i$ is the average value of the activity $s_i(t)$ at node $i$ and $\sigma_{i}$ is the standard deviations \cite{dayan}.

We calculated the mutual information (fig. \ref{fig.mi}) and correlation (fig. \ref{fig.corr}) in this way for all pair of nodes and the same set of parameters as for the similarity ($P_{EE} = 0.1, T=5.2, P_{QE}=0.15, 0.263$ and $0.4$).

To determine the value of spurious correlations, we also computed $C(s_i,s_j,\tau)$ using $s_i$ and $s_j$ taken from different realizations with the same parameter values. Since these signals are perforce independent, the maximum value reached sets a threshold below which correlations in the systems cannot be distinguished from spurious correlation. 

\begin{figure}[ht!]
 \centering
\includegraphics[width=.6\linewidth]{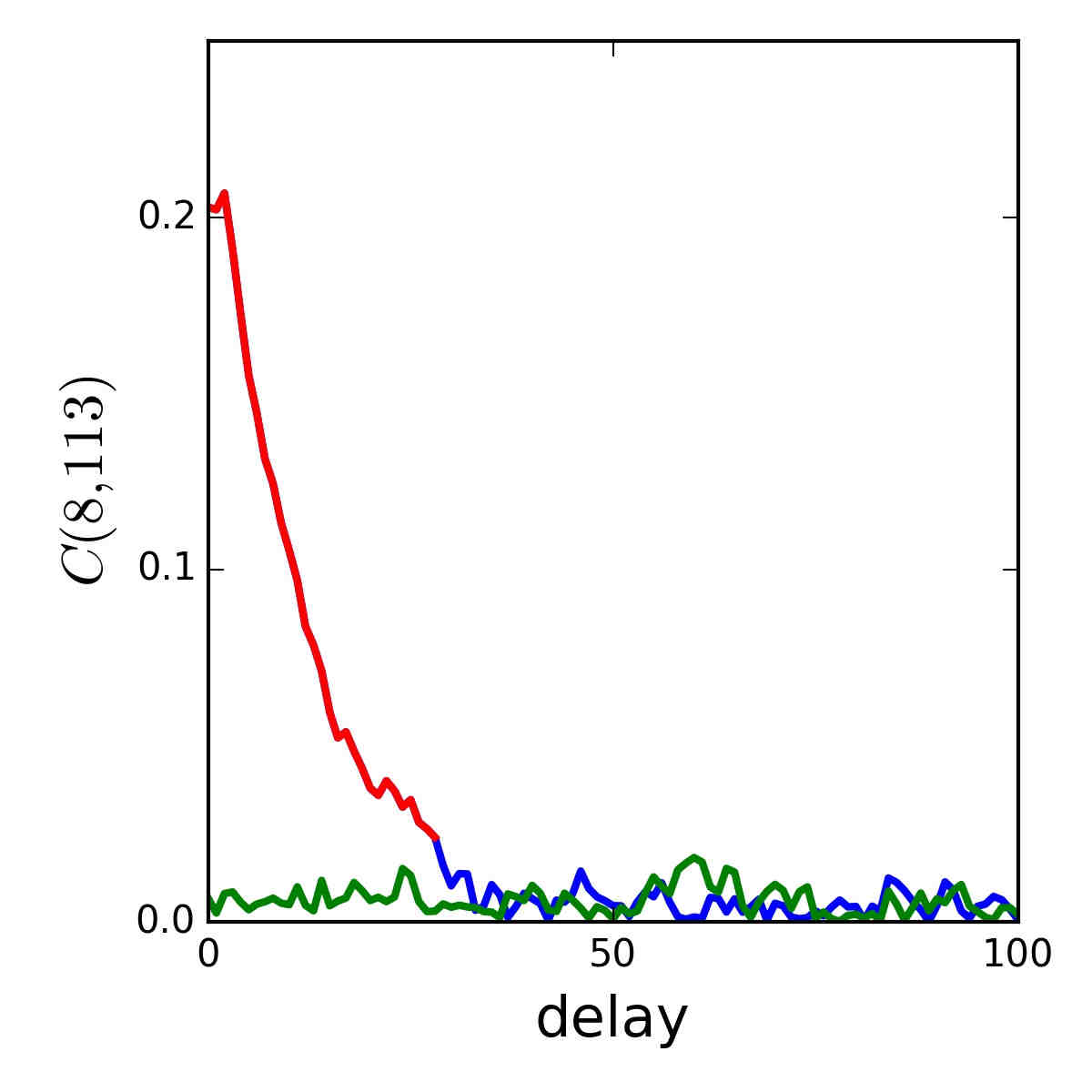}
\caption{\footnotesize{The correlation between a pair of nodes from different realizations (green curve) should be zero due to the independence, so it establish a threshold to consider a correlation as spurious. The time the nodes were correlated (red dots) is determined when the correlation obtained for nodes from the same realization (blue curve) is lower than the threshold.}}
\label{fig.corrExample}
\end{figure}

\begin{figure}[ht!]
 \centering
\includegraphics[width=\linewidth]{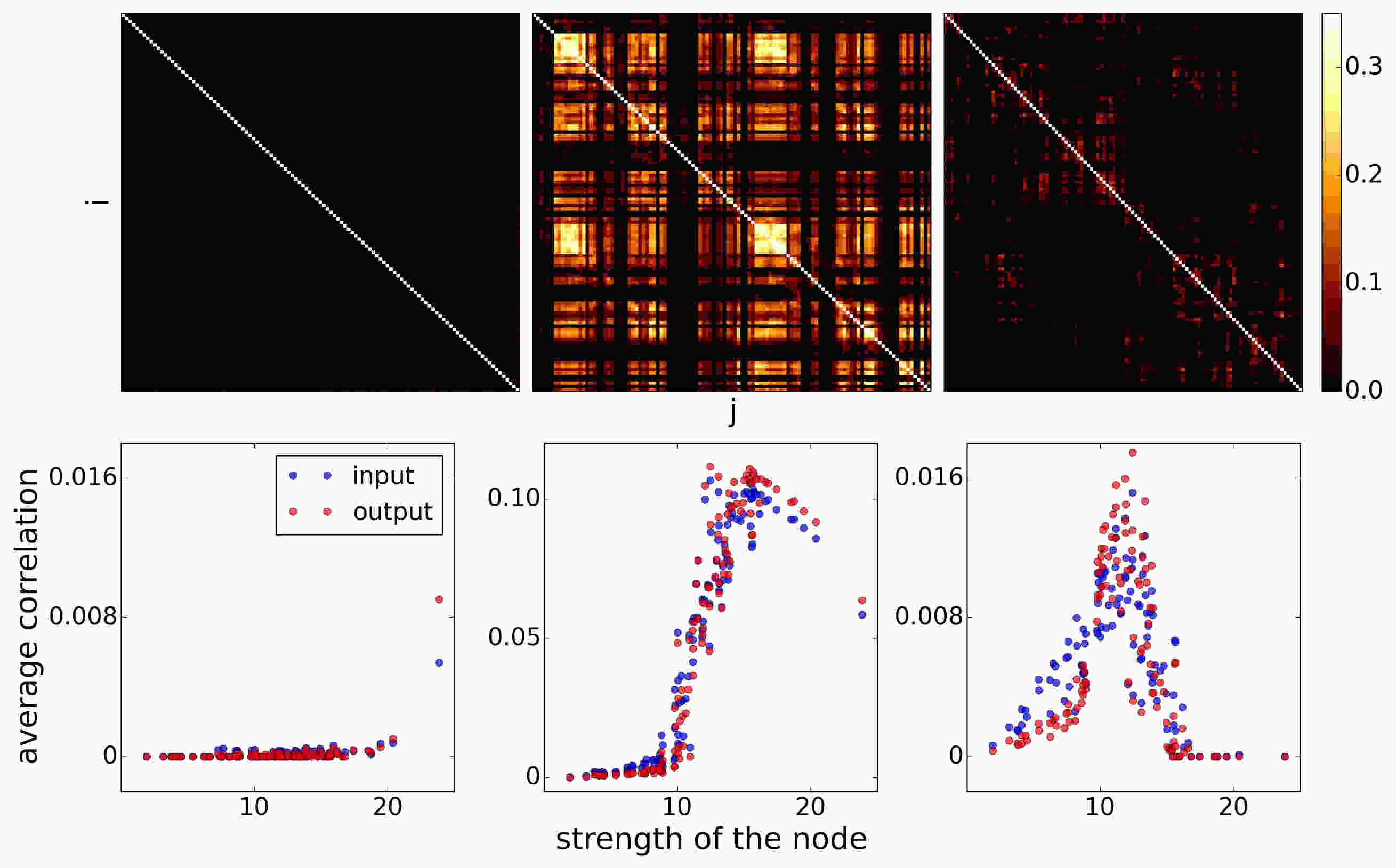}
\caption{\footnotesize{Top: maximum correlation between pairs of nodes with $P_{EE} = 0.1, T = 5.2$ and $P_{QE}=0.15, 0.263$ and $0.4$. Bottom: average maximum correlation over inputs (blue) or outputs (red) as a function of the strength of the nodes.}}
\label{fig.corr}
\end{figure}

\begin{figure}[ht!]
 \centering
\includegraphics[width=\linewidth]{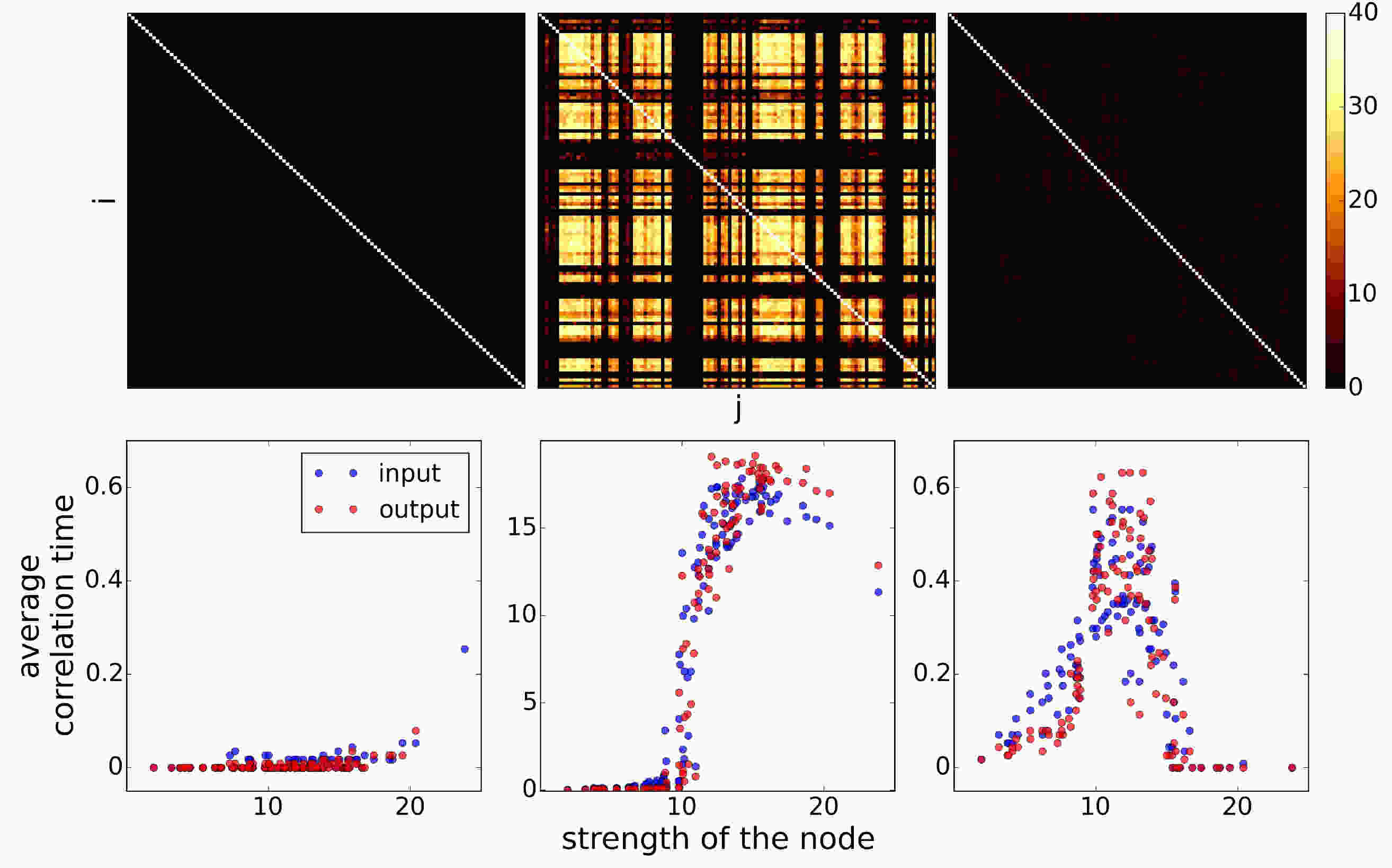}
\caption{\footnotesize{Top: time interval over which the nodes are correlated with $P_{EE} = 0.1, T = 5.2$ and $P_{QE}=0.15, 0.263$ and $0.4$. Bottom: average correlation time over inputs (blue) or outputs (red) as a function of the strength of the nodes.}}
\label{fig.corr_iTau}
\end{figure}

\begin{figure}[ht!]
 \centering
\includegraphics[width=\linewidth]{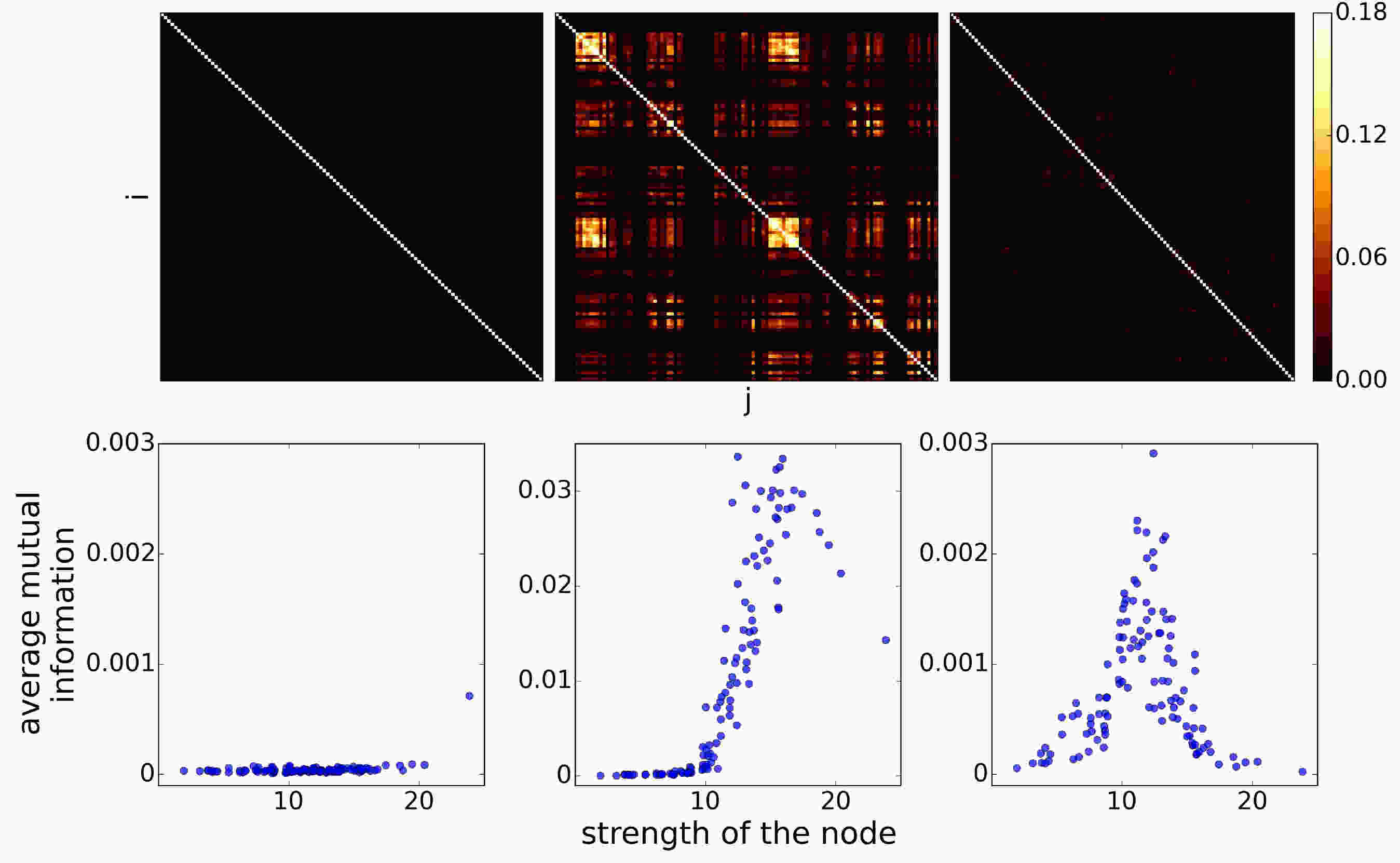}
\caption{\footnotesize{Top: maximum mutual information between pairs of nodes with $P_{EE} = 0.1, T = 5.2$ and $P_{QE}=0.15, 0.263$ and $0.4$. Bottom: average maximum mutual information over inputs as a function of the strength of the nodes.}}
\label{fig.mi}
\end{figure}

\begin{figure}[ht!]
 \centering
\includegraphics[width=\linewidth]{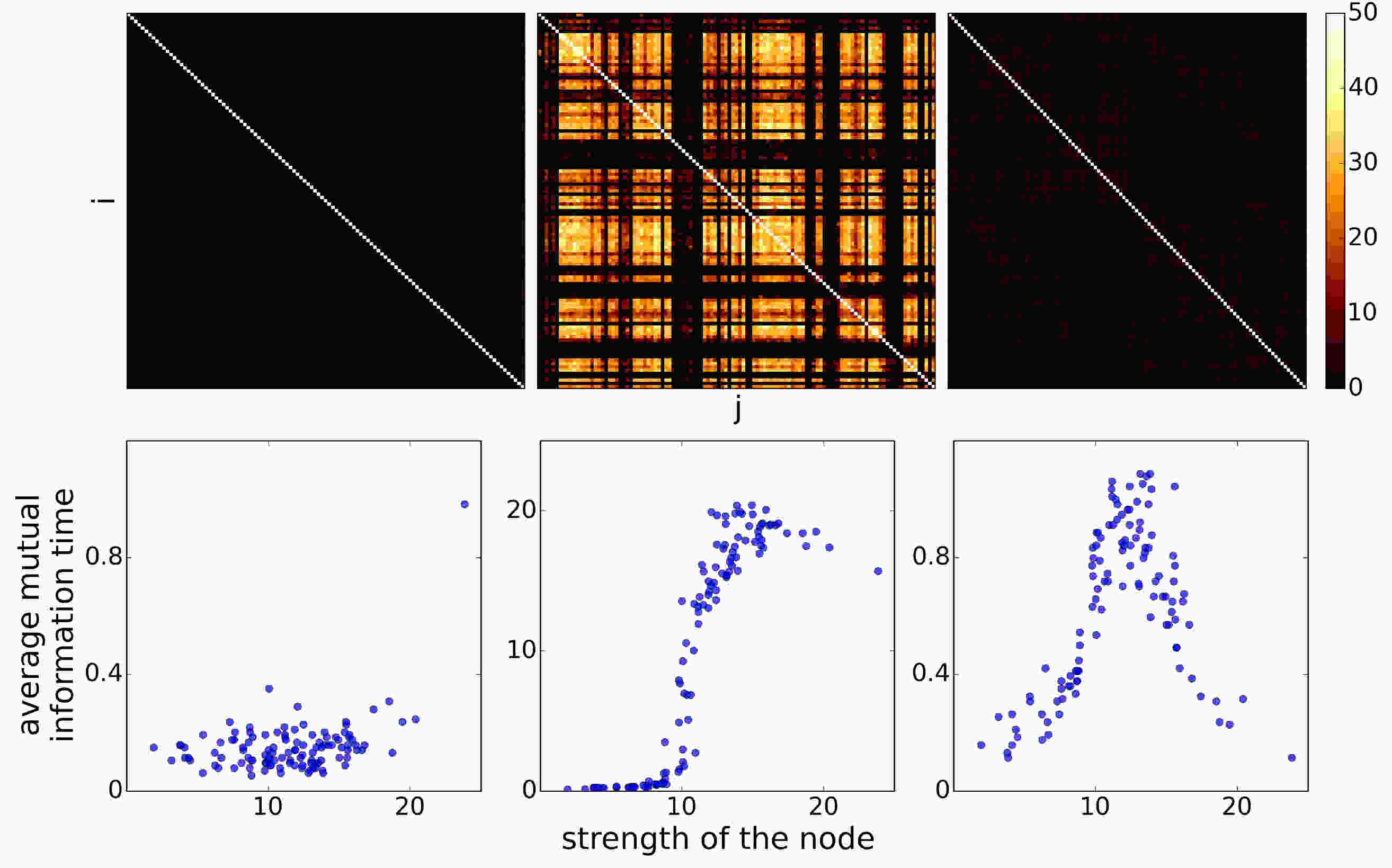}
\caption{\footnotesize{Top: time interval over which the nodes have a mutual information larger than the one obtained for nodes from different realizations. For a system with $P_{EE} = 0.1, T = 5.2$ and $P_{QE}=0.15, 0.263$ and $0.4$. Bottom: average mutual information time over inputs as a function of the strength of the nodes.}}
\label{fig.mi_iTau}
\end{figure}

When we compute the actual correlation between nodes from the same realization, we define the correlation time as the first time the value of the correlation falls below the threshold established above  (fig. \ref{fig.corrExample}). We record the maximum correlation and mutual information (figures \ref{fig.corr}, \ref{fig.mi}) between the pair of nodes and the time during which the nodes were correlated or sharing information (figures \ref{fig.corr_iTau}, \ref{fig.mi_iTau}). 

\cleardoublepage

\section*{Acknowledgements}

B.V-R. knowledges CONACyT for the scholarship during her graduated studies. B.V-R. and H.L. knowledge DGAPA for the partial support through projects PAPIIT IN110016. B.V-R. and H.L. acknowledge useful early discussion with professor Dante Chialvo. The MRI acquisition was supported by Centre d'Imagerie BioMédicale (CIBM) of the UNIL, UNIGE, HUG, CHUV, EPFL, and the Leenaards and Jeantet Foundations in Switzerland as well as Swiss National Science Foundation $310030_156874$.

\section*{Author contributions statement}

B.V-R and H.L designed the experiments. B.V-R performed the numerical simulations. B.V-R, A.A-K and H.L analyzed the results. A.G and P.H provided the connectome data. B.V-R, A.A-K, O.S and H.L contributed and reviewed the manuscript.

\end{document}